\documentclass[twocolumn, pdflatex, sn-mathphys-num]{sn-jnl}

\usepackage{float}

\newgeometry{margin=1in, twocolumn}
\usepackage{hyperref}
\usepackage{graphicx}
\usepackage{multirow}
\usepackage{amsmath,amssymb,amsfonts}
\usepackage{amsthm}
\usepackage{mathrsfs}
\usepackage[title]{appendix}
\usepackage{xcolor}
\usepackage{textcomp}
\usepackage{manyfoot}
\usepackage{booktabs}
\usepackage{siunitx}
\usepackage{algorithm}
\usepackage{algorithmicx}
\usepackage{algpseudocode}
\usepackage{listings}
\usepackage{url}
\usepackage{breakurl} 
\def\UrlBreaks{\do\/\do-\do_\do.\do,\do:\do\do\_;}

\theoremstyle{thmstyleone}

\theoremstyle{thmstyletwo}

\theoremstyle{thmstylethree}

\begin{document}
\def\UrlBreaks{\do\/\do-\do_\do\.}

\title[Article Title]{Ultra-Low-Loss Silicon Nitride on Sapphire for Broad-Transparency Nonlinear and Quantum Photonics}

\author*[1]{\fnm{Abdur-Raheem} \sur{Al-Hallak}}\email{abhallak@umich.edu}
\equalcont{These authors contributed equally to this work.}

\author[1]{\fnm{Shuai} \sur{Liu}}
\equalcont{These authors contributed equally to this work.}

\author[1]{\fnm{Kailu} \sur{Zhou}}
\equalcont{These authors contributed equally to this work.}

\author[1]{\fnm{Jiangnan} \sur{Liu}}

\author[1]{\fnm{Shawn} \sur{Chen}}

\author[1]{\fnm{James} \sur{Hu}}

\author[1]{\fnm{Ruhi} \sur{Yusuf}}

\author[1]{\fnm{Christopher} \sur{Rodriguez}}

\author[1]{\fnm{Maya} \sur{Sarram}}

\author[1]{\fnm{Yiming} \sur{Lang}}

\author*[1, 2]{\fnm{Zetian} \sur{Mi}}\email{ztmi@umich.edu}

\author*[1, 2]{\fnm{Zheshen} \sur{Zhang}}\email{zzsh@umich.edu}

\affil[1]{\orgdiv{Department of Electrical Engineering and Computer Science}, \orgname{University of Michigan}, \orgaddress{\street{1301 Beal Avenue}, \city{Ann Arbor}, \postcode{48109}, \state{MI}, \country{USA}}}

\affil[2]{\orgdiv{Quantum Research Institute}, \orgname{University of Michigan}, \orgaddress{\city{Ann Arbor}, \postcode{48109}, \state{MI}, \country{USA}}}

\abstract{The field of photonic integrated circuits (PIC) has flourished in the past two decades, fueling numerous cutting-edge applications across sensing, networking, data interconnect, and quantum information processing. As a guiding material for PIC, Si$_3$N$_4$ has seen extensive use for its ultra-low loss, broad transparency, and diversity in implementation across both thin and thick films. Although the standard, traditional silicon dioxide (SiO$_2$) on silicon (Si) substrates that underpin the majority of Si$_3$N$_4$ photonics face drawbacks in the form of long-wavelength transparency limited by SiO$_2$, high-stress deposition for anomalous dispersion thick-film Si$_3$N$_4$, and leakage loss to the Si layer for low-confinement thin-film Si$_3$N$_4$. Featuring increased long-wavelength transparency into the mid-infrared, low-stress deposition of Si$_3$N$_4$, and a low index, this work investigates sapphire substrates as alternate hosts for Si$_3$N$_4$ photonics with greater spectral coverage and reduced fabrication complexity. This work presents a robust method of fabricating ultra-low loss photonic integrated circuits on a 500-nm-thick Si$_3$N$_4$-on-sapphire platform, exhibiting record-low losses below $0.1 \rm \;dB/cm$. Implemented using this process are high-Q microrings with intrinsic quality factors in excess of $4.5\times10^6$ and coupled-ring photonic molecules to support nonlinear gain. Leveraging the achievable low loss and high-Q, this work further reports the first demonstration of Kerr-comb and soliton generation on the Si$_3$N$_4$-on-sapphire platform. These advances in loss, quality factor, and soliton generation on this versatile, broad-transparency platform pave the way for future work in spectroscopy and quantum-enhanced sensing across previously prohibited spectral regions for Si$_3$N$_4$ photonics with reduced fabrication complexity.} 

\keywords{Silicon nitride, sapphire, ultra-low loss, frequency comb, soliton}

\maketitle

\section{Introduction}\label{sec1}

For over a decade, the integrated photonics community has witnessed the rise of silicon nitride (Si$_3$N$_4$) as the preeminent material platform for low-loss and nonlinear integrated photonics. To date, Si$_3$N$_4$ remains the lowest-loss material platform for planar waveguides near the telecom band. High quality films of this wide bandgap material permit the fabrication of integrated planar waveguides with loss as low as 0.034 dB/m in its high-aspect ratio, low-confinement guise \cite{SiN720M}. These ultra low losses underpin on-chip microring cavities with quality factors as high as $7.2 \times 10^8$, enabling low-threshold, narrow-linewidth Brillouin lasers \cite{SiN720M, SubHtzBrillouin18, VisIntegBrillouin21, PhotMolBrillouin24, LargeModeVolBrillouin25, OctSpanIntegBrillouin26}, ultra-low-noise external cavity lasers \cite{HzLineWidthECLSIN21, HighPerfLasersForSiN21, WidelyTuneSiNLaserLipson23, Sub100HzSiNLaser24, LowLWCoilStabIsoFreeSiNLaser25}, and high-quality reference cavities \cite{OctaveSpanRefCavity26}. The broad transparency of Si$_3$N$_4$ allows this superlative performance to persist at shorter wavelengths, where intrinsic quality factors above $6\times10^6$ are still achievable for thin Si$_3$N$_4$ resonators operating in the 400-500 nm band \cite{ThinSi3N4UVHighQ2021, ULLThinSi3N4Vis22}. When loss is of the utmost importance, the aforementioned low-confinement, thin Si$_3$N$_4$ configuration is preferred; however, high-confinement, thick Si$_3$N$_4$ has similarly seen great use, offering denser integration---with minimum bending radii on the scale of tens of microns, rather than millimeter-scale---and anomalous dispersion to facilitate broadband phase-matching for nonlinear gain \cite{BowersSiNRev22}. While high-confinement implementations do see increased loss in comparison to low-confinement, intrinsic quality factors up to 67 million have been demonstrated near 1550 nm, corresponding to losses as low as 0.43 dB/m \cite{LipsonSiN67MQ17}. The high intrinsic quality factors and anomalous dispersion accessible on the thick Si$_3$N$_4$ platform and the appreciable third-order nonlinearity of the material ($n_2 \approx 2.4\times10^{-19}\rm m^2/W$ \cite{ThermalAndKerrSiN08}) give rise to $\chi^{(3)}$-based low-threshold parametric oscillators with the ability to generate a diverse array of nonlinear states. This combination of properties has made Si$_3$N$_4$ the premier platform for the generation and investigation of dissipative Kerr solitons (DKS), with hallmark works demonstrating octave-spanning \cite{OctSpanSoliton2017}, laser-integrated \cite{IntTurnkeySoliton2020}, deterministic single-\cite{CuFreeDeterSoliton25}, and high-conversion efficiency solitons \cite{SurpNLOConvEffSoliton23}, among a multitude of other advances. In configurations without optical cavities, the same low-loss, nonlinearity, and dispersion-engineering features have also been leveraged to create broadband parametric amplifiers \cite{Si3N4ParaAmpNat2022, Si3N4BroadParaAmpNat2025} alongside devices for supercontinuum generation \cite{Si3N4SuperCont26}. Further leveraging these superlative qualities, the exceptional low loss in particular has allowed Si$_3$N$_4$ to go beyond classical applications and generate exotic quantum squeezed states of light, providing a basis for quantum-enhanced sensing, quantum communication, and quantum computing \cite{SQZNanoPhotMolSi3N42021, StrongSQZBeyond3p5dB2025, QuadSQZNanoPhotRes2025, IntDifSQZSi3N42026, SQZThinFilmShuai2026}.

Of all the groundbreaking work in Si$_3$N$_4$ photonics, the vast majority has been focusing on the platform based on low-pressure chemical vapor deposition (LPCVD) stoichiometric Si$_3$N$_4$ films on SiO$_2$-on-Si substrates by virtue of its low loss, good index contrast, and ease of integration with the standard silicon-based foundry lines. However, this mainstream Si$_3$N$_4$ platform exhibits high stress, yielding cracking throughout the wafer beyond 400 nm in thickness \cite{LipsonSiNRev21, BowersSiNRev22}. Yet, 600-nm deposition is needed to enter the anomalous dispersion regime to support a variety of nonlinear and quantum photonics applications. Efforts to mitigate cracking or crack propagation in thick-film Si$_3$N$_4$, including multi-step deposition \cite{MultiStepRot10M19}, the photonic damascene process \cite{PhotDamasSiN30M21}, and subtractive methods \cite{LipsonSiN67MQ17, ShuaiACS25}, have been pursued to achieve high yield for wafer-scale low-loss photonic integrated circuits. On the other end of the spectrum, thin Si$_3$N$_4$ films, albeit free from stress-induced cracking, are plagued by increased mode size and reduced confinement that result in leakage loss to the Si substrate. Consequently, thermal SiO$_2$ layers of up to 10-15 $\mu$m in thickness are required for low-confinement Si$_3$N$_4$ devices \cite{LipsonSiNRev21, Si3N4RevIlya24}, necessitating hundreds of hours of growth time that greatly increases fabrication complexity and the manufacturing cost of base wafer. Lastly, a common drawback to both thin and thick-film Si$_3$N$_4$ on SiO$_2$ platforms is the long-wavelength opacity of SiO$_2$ that limits the overall transparency window to $\sim$4 $\mu$m \cite{MIRCommentary2010, MIRSiPhotonRev24}, even though Si$_3$N$_4$ itself possesses a broad transparency window (up to $\approx$ 6.7 $\mu$m) that stretches well into the mid-infrared (MIR).

To address the aforementioned challenges for SiO$_2$-on-Si substrates, crystalline aluminum oxide (Al$_2$O$_3$), i.e., sapphire, has emerged as an intriguing candidate substrate for Si$_3$N$_4$ photonics. Sapphire's transparency into MIR wavelengths exceeds that of SiO$_2$ by over a micron \cite{MIRCommentary2010, IntPhotMatforMIRRev2020, MIRSiPhotonRev24}, unlocking the long-wavelength potential of Si$_3$N$_4$ photonics, as previously achieved in sapphire on silicon photonics that extends low-loss operation out to 5.5 $\mu$m \cite{SOSForMIR2010, LowPropSOSforMIR2011, SOSHighQMIR2013, SiMIR5p5micron2010} in rigid devices without involving suspending the waveguide \cite{IntPhotMatforMIRRev2020}. Moreover, since the entire sapphire wafer exhibits a refractive index smaller than Si$_3$N$_4$ (1.75 vs 1.99 at 1550 nm), both high and low confinement waveguides can be fabricated on the same base-wafer, eliminating added complexity and cost arising from differing thicknesses of thermal-oxide. Lastly, sapphire inherently supports direct crack-free, low-stress deposition of LPCVD Si$_3$N$_4$ in excess of 1300 nm in thickness \cite{ThickSiliconNitrideOnSapphire24}, vastly reducing the fabrication complexity in comparison to Si substrates.

\begin{figure*}
\centering\includegraphics[width=1\linewidth]{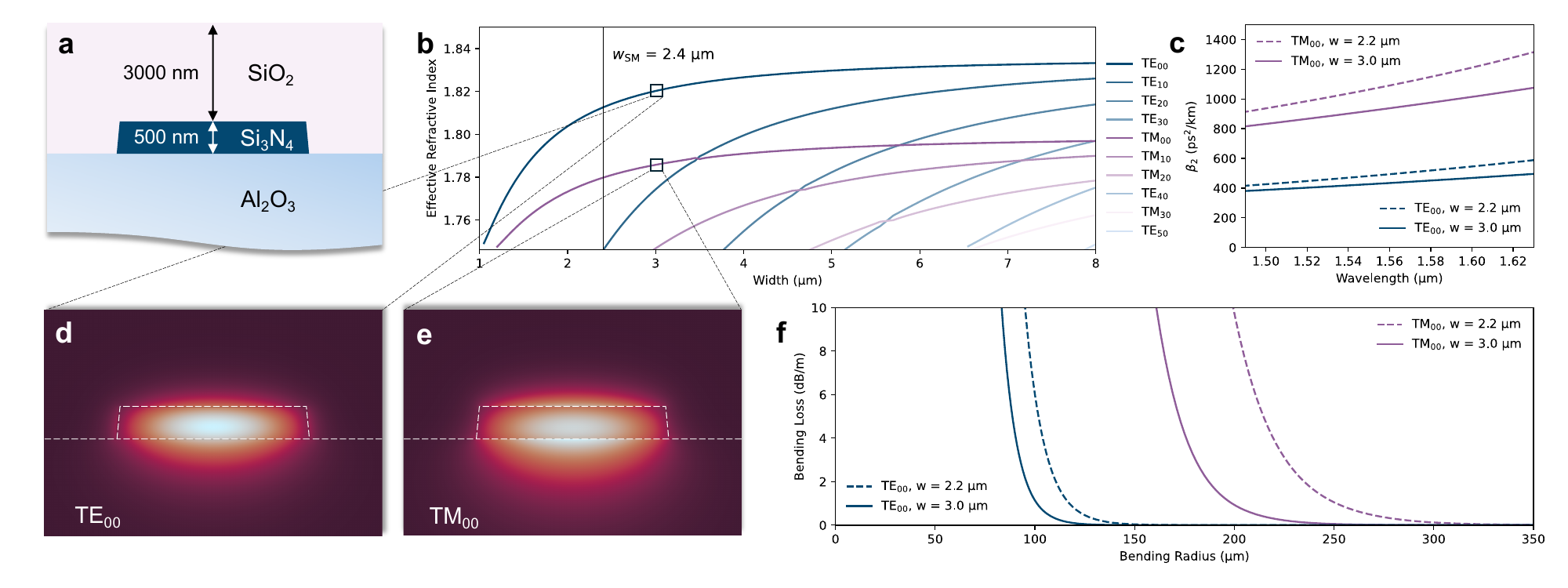}
\caption{{\bf Simulation and Design.} \textbf{a,} Geometry of the designed Si$_3$N$_4$ on sapphire platform. \textbf{b,} Simulated TE and TM mode families and their respective effective refractive indices for the geometry shown in (\textbf{a}) as a function of width at 1550 nm, indicating single-mode width of 2.4 $\mu$m. \textbf{c,} Simulated group velocity dispersion ($\beta_2$) of the fundamental TE$_{00}$ and TM$_{00}$ modes for widths of 2.2 and 3.0 $\mu$m and wavelengths between 1490 and 1630 nm. Simulated transverse mode profiles of the fundamental TE$_{00}$ (\textbf{d}) and TM$_{00}$ (\textbf{e}) modes for a width of 3.0 $\mu$m and a wavelength of 1550 nm. \textbf{f,} Simulated bending losses as a function of bending radius at 1550 nm for both fundamental TE$_{00}$ and TM$_{00}$ modes at widths of 2.2 and 3.0 $\mu$m. All simulated data and mode-profiles are computed using COMSOL Multiphysics.}
\label{fig:sim}
\end{figure*}

In addition to the solutions sapphire provides to existing obstacles, sapphire substrates are home to their own eco-system of mature photonic and electronic components that provide new opportunities for integration. In particular, sapphire substrates act as a host for the epitaxial growth of III-nitrides, allowing for future monolithic integration with III-nitride gain media, with bandgaps tunable across the entire visible spectrum and beyond \cite{ThreeNitrideRev2018}, or $\chi^{(2)}$ and piezoelectric media \cite{AlNPhotRev23, MehediPiezoScAlN25}. Titanium (Ti) doping also provides an additional degree of freedom, allowing the substrate itself to act as a gain medium for broadband lasing in the visible and near-visible band centered around 800 nm. Pioneering work by Wang {\em et al.} first explored this prospect of an integrated Ti:sapphire-Si$_3$N$_4$ laser through a bonded sapphire-Si$_3$N$_4$-sapphire approach, resulting in losses as low as 12 dB/m in the visible and lasing from 730 nm to 830 nm \cite{SOSTiSapphHongTang23, SOSHongTangVis24}. These numerous possibilities for integration open the doors for heterogeneous, multi-functional photonic platforms based on Si$_3$N$_4$-on-Sapphire. 

With the convergence of these enticing advantages at-hand, we fabricate photonic integrated circuits in a 500 nm Si$_3$N$_4$-on-sapphire platform, achieving intrinsic quality factors up to $4.6 \times10^6$, corresponding to an ultra-low (sub 10 dB/m \cite{ULLThinSi3N4Vis22}) loss of 7.5 dB/m in the telecommunication window. We further showcase the rich nonlinear dynamics in Si$_3$N$_4$-on-sapphire photonics by demonstrating Kerr-comb and soliton generation for the first time on this versatile platform, paving the way for future development of this platform for sensing, communication, spectroscopic, and computing applications.

\begin{figure*}
\centering\includegraphics[width=1\linewidth]{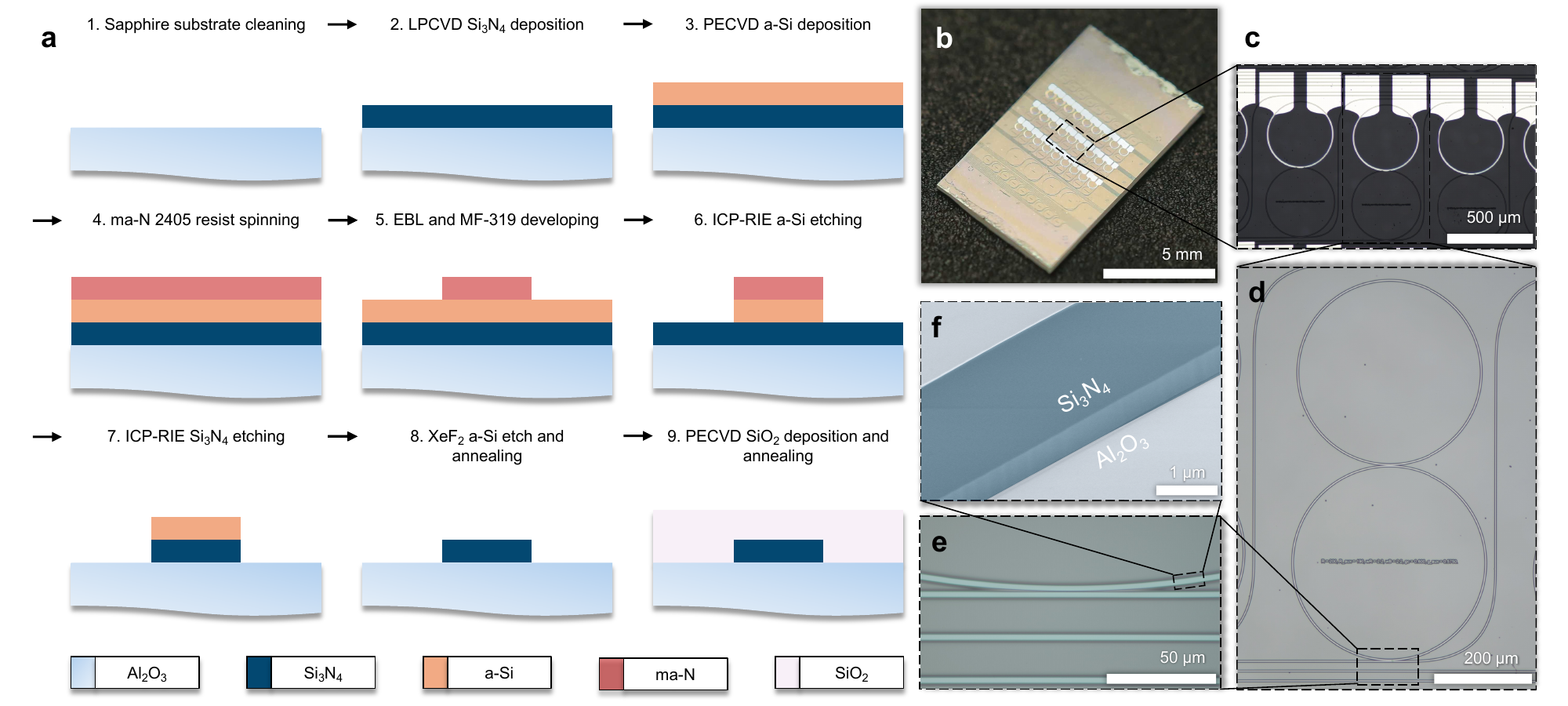}
\caption{{\bf Device Fabrication.} \textbf{a,} Overview of fabrication flow for a Si$_3$N$_4$-on-sapphire PIC. \textbf{b,} Picture of a complete Si$_3$N$_4$-on-sapphire chip. Optical microscope images of an array of coupled microrings photonic molecules with microheaters at 5X magnification (\textbf{c}), a single coupled microring without microheaters at 10X magnification (\textbf{d}), and the bus-resonator coupling region for a coupled microring at 100X magnification (\textbf{e}). \textbf{f,} False colored scanning electron microscope image of an exposed Si$_3$N$_4$ waveguide on a sapphire substrate (scale indicated).}
\label{fig:fab}
\end{figure*}

\section{Results}
\subsection{Device Geometry and Simulation}
Figure \ref{fig:sim}\textbf{a} shows the designed geometry for our Si$_3$N$_4$-on-sapphire device: a thickness of 500 nm is chosen for the Si$_3$N$_4$ waveguide. At this thickness, the waveguide operates in the low-to-moderate confinement regime in contrast to moderate-to-high confinement in the conventional Si$_3$N$_4$-on-SiO$_2$ structure due to the increased index of sapphire in comparison to SiO$_2$ (1.75 vs 1.44 at 1550 nm). This reduced-confinement behavior becomes clear in Fig. \ref{fig:sim}\textbf{b}, where the available mode families with respect to waveguide width are plotted, showing a large single-mode width of 2.4 $\mu$m at a wavelength of 1550 nm. Figure \ref{fig:sim}\textbf{c} further depicts the simulated group velocity dispersion for waveguide widths of 2.2 and 3.0 $\mu$m, demonstrating strong normal dispersion ($\beta_2 > 0$) across our wavelength range of interest (1490 to 1630 nm). The normal dispersion precludes broadband phase matching for nonlinear gain, motivating our later use of coupled-microring photonic molecules to facilitate Kerr comb generation. Apart from the thickness of the Si$_3$N$_4$ layer, the SiO$_2$ cladding has a significant impact on the mode-confinement of quasi-TE and TM modes. Owing to the reduced index of SiO$_2$ in comparison to sapphire, the out-of-plane boundary condition for the waveguide becomes asymmetric, leading to very different confinement for quasi-TE and TM modes, as visually identified in Figs. \ref{fig:sim}\textbf{d} and \textbf{e}. With TE$_{00}$ being polarized in-plane, the out-of-plane asymmetry does not significantly impact its confinement, with the mode well confined in the guide region. In contrast, the asymmetry pushes the out-of-plane-polarized TM$_{00}$ mode towards the higher index sapphire substrate, resulting in significantly more confinement within the sapphire region than TE$_{00}$. This effect further manifests in Fig. \ref{fig:sim}\textbf{f} as the disparity in bending loss between the two polarizations. While the TE$_{00}$ mode exhibits negligible bending loss for both 2.2 and 3.0 $\mu$m wide waveguides to below a radius of 150 $\mu$m at 1550 nm, bending loss for the TM$_{00}$ mode becomes substantial for radii below 300 $\mu$m, due to leakage of the mode into the sapphire substrate. With the sharp contrast in the bending loss for the two mode families observed, we choose to focused on the use of TE-polarization for the more compact structures and include both 200 and 400 $\mu$m radius microring resonators in our final chip design.

\subsection{Silicon Nitride on Sapphire Device Fabrication}
The fabrication of ultra-low-loss PICs on the Si$_3$N$_4$-on-sapphire platform builds on the amorphous silicon (a-Si) hardmask etching process first reported by Liu {\em et al.} in Ref. \citenum{ShuaiACS25}. The fabrication flow is outlined in Fig. \ref{fig:fab}\textbf{a} and detailed in Methods \ref{sec:FabMethods}. The fabrication of the complete devices shown in Fig. \ref{fig:fab}\textbf{b}-\textbf{e} begins by depositing a low-stress 500 nm thick layer of LPCVD Si$_3$N$_4$ on a c-plane sapphire wafer, exceeding the thickness ($< 400$ nm) that can be reliably deposited on traditional SiO$_2$-on-Si wafers without suffering from cracking \cite{ThickSiliconNitrideOnSapphire24}. A layer of PECVD a-Si is subsequently deposited for high-quality, near vertical etching to define a patterned hardmask that can produce the smooth waveguide sidewalls (Fig. \ref{fig:fab}\textbf{f}) needed to suppress Rayleigh-type scattering loss. Patterning of the hardmask is achieved through electron-beam lithography and negative ma-N 2405 electron beam resist, enabling accurate definition of coupling gaps down to 500 nm in width. To suppress material absorption loss, two rounds of high-temperature (1100 $^\circ$C) nitrogen annealing are performed for 6 hours to drive out hydrogen and reduce its induced loss near 1520 nm. The first round is launched after the Si$_3$N$_4$ layer is fully patterned, while the second round comes after the deposition of the 3 $\mu$m PECVD SiO$_2$ cladding to further densify the cladding layer. Upon completing the photonic layer after the final annealing step, the process proceeds to the fabrication of microheaters (Fig. \ref{fig:fab}\textbf{b} and \textbf{c}) necessary to control the splitting of the photonic molecules. The microheaters are defined through platinum deposition and a standard photolithography-liftoff process. Finally, the chip facets are diced, polished, and cleaned to reveal edge couplers, permitting testing to commence.

\subsection{Ultra-low-Loss Si$_3$N$_4$ on Sapphire Microrings}
\label{sec:multimodeperf}
\begin{figure*}
\centering\includegraphics[width=1\linewidth]{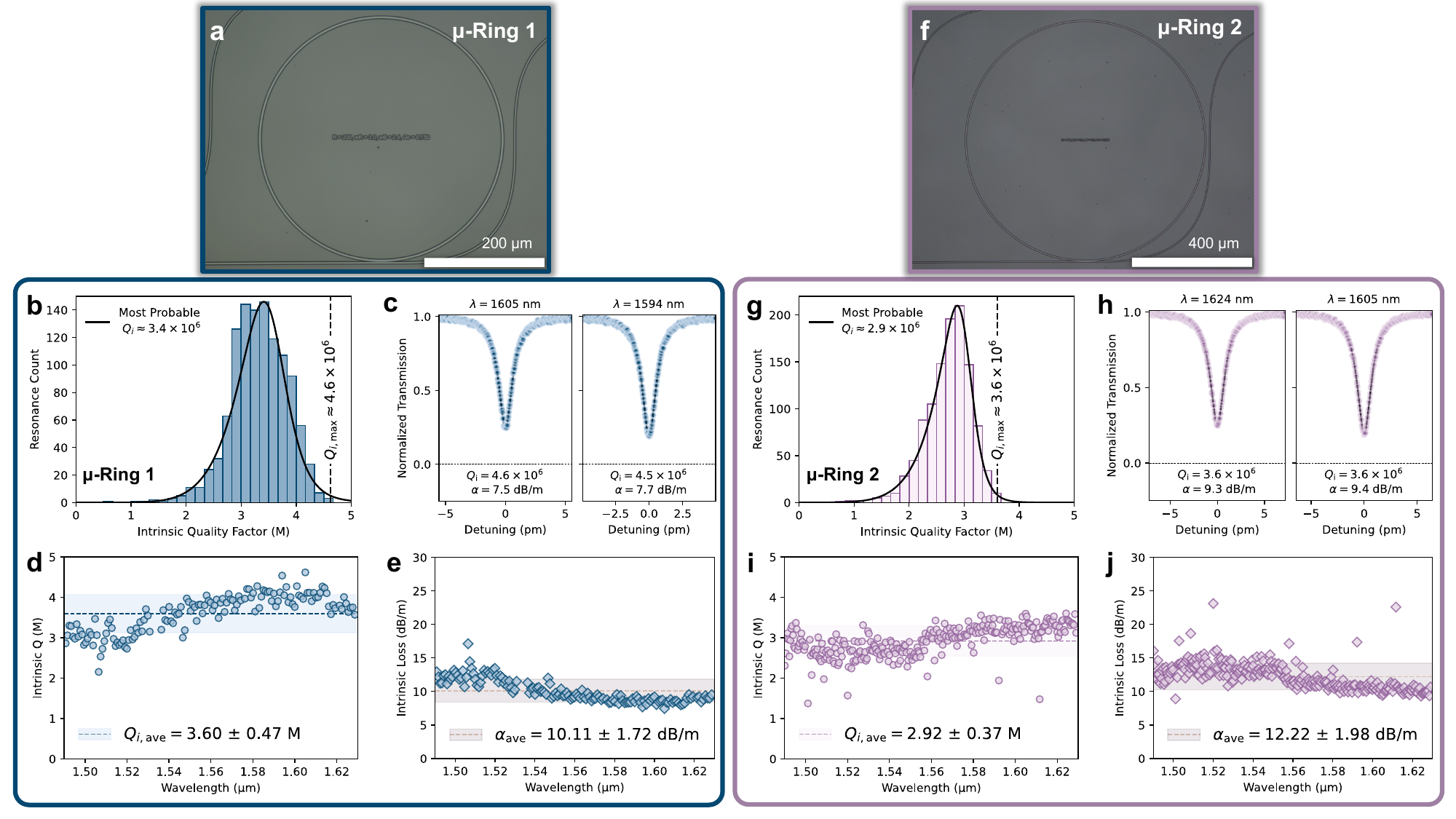}
\caption{{\bf Loss Characterization of Multi-Mode Microrings.} Optical microscope images of multi-mode (3 $\mu$m width) microring resonators with radii of 200 $\mu$m (\textbf{a}) and 400 $\mu$m (\textbf{f}) with scales indicated. \textbf{b,} Distribution of extracted intrinsic-Q ($Q_i$) for TE polarization across 6 type-1 micorings between 1490 and 1630 nm. Burr fitting of the distribution indicates a most-probable $Q_i$ of $3.4\times10^6$. \textbf{c,} Fitted transmission for the two highest Q resonances in the highest performance resonator, showing intrinsic quality factors (losses) of $4.6\times10^6$ (7.5 dB/m) and $4.5\times10^6$ (7.7 dB/m) respectively. Trends of intrinsic-Q (\textbf{d}) and loss (\textbf{e}) from 1490 nm to 1630 nm for TE-polarization in the highest performance type 1 microring with averages of $3.6\pm0.47$ M and $10.11\pm1.72$ dB/m indicated. Shaded regions indicate a 1-sigma deviation from the mean. Distribution of intrinsic-Q for TM (\textbf{g}) polarization across 4 type-2 micorings between 1490 and 1630 nm, with a most-probable $Q_i$ indicated. Fitted transmission for the two highest-Q resonances in the highest performance resonator for and TM (\textbf{h}) polarization. Trends of intrinsic-Q (\textbf{i}) and loss (\textbf{j}) from 1490 nm to 1630 nm for TM-polarization in the highest performance type-2 microring with averages of $2.92\pm0.37$ M and $12.22\pm1.98$ dB/m indicated.} 
\label{fig:Multimode}
\end{figure*}

\begin{figure*}
\centering\includegraphics[width=1\linewidth]{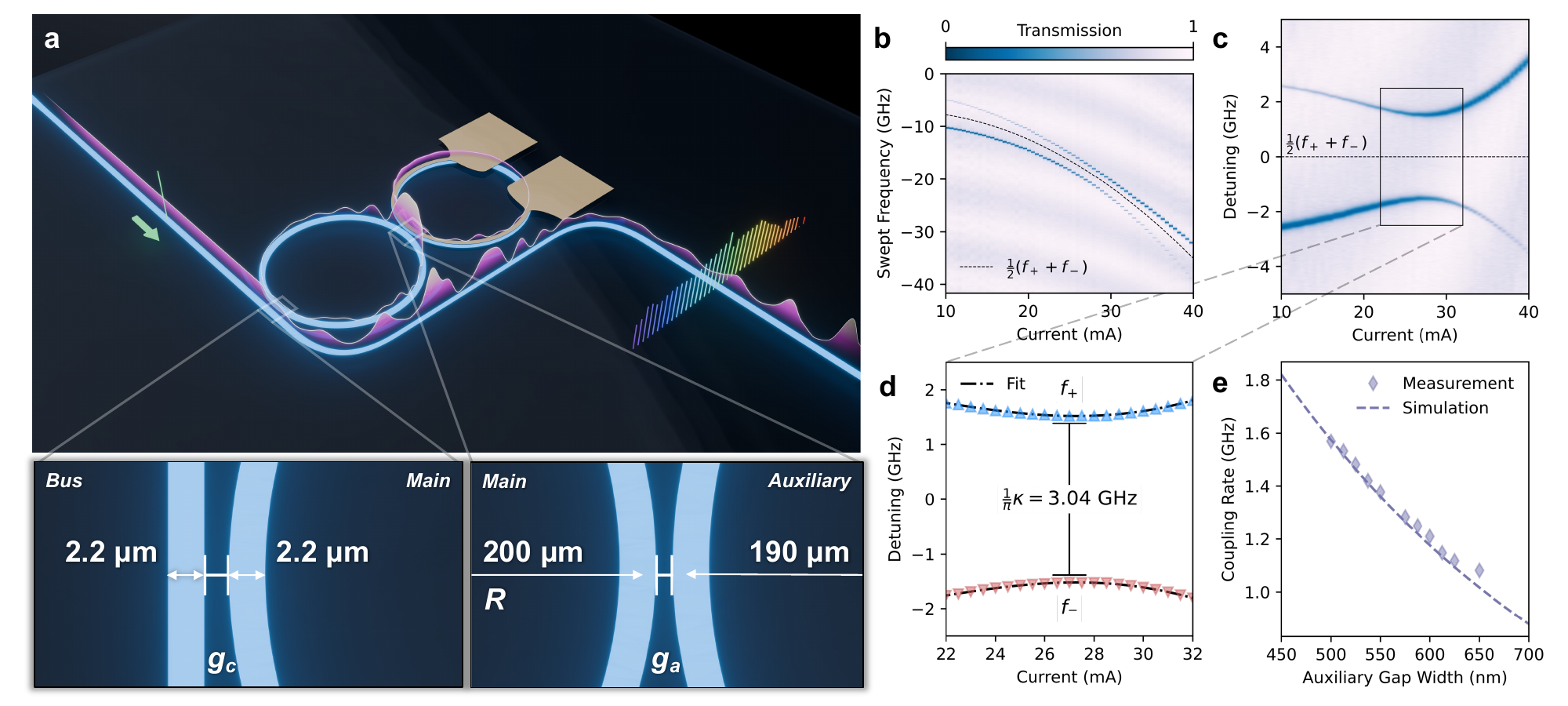}
\caption{{\bf Coupled Microring Photonic Molecule.} \textbf{a,} Diagram of the coupled-microring photonic molecule. \textbf{b,} Recorded transmission spectra of a photonic-molecule split resonance near 1575 nm across a range of applied heater currents. Centerline indicates common-mode movement of the symmetric and anti-symmetric modes $f_\pm$. \textbf{c,} Transmission spectra across applied heater currents with common-mode movement of $f_\pm$ removed and detuning from the centerline calibrated by a fiber-loop resonator reference cavity. \textbf{d,} Zoomed-in view of avoided mode crossing behavior of the split resonances with minimum splitting of $\kappa/\pi = 3.04 \rm \; GHz$ indicated. \textbf{e,} Coupling rates $\kappa/2\pi$ near 1575 nm extracted from the measured avoided mode crossings for coupled-microring photonic molecules with main-auxiliary microring gaps ranging from 500 to 650 nm. Simulated coupling rates overlaid.} 
\label{fig:Coupledring}
\end{figure*}

The intrinsic loss of the Si$_3$N$_4$-on-sapphire platform is characterized by the transmission spectra of the fabricated microring resonators across 1490 nm to 1630 nm using a tunable laser and low-noise photodetector. The intrinsic-Q is then extracted from the loaded-Q measured from the Lorentzian-fitted transmission spectra. Figure \ref{fig:Multimode} showcases the ultra-low-loss performance of the multimode (3 $\mu$m wide) microring resonators for both TE- and TM-polarized modes in 200 $\mu$m (type-1) and 400 $\mu$m (type-2) radius microrings, shown in Fig. \ref{fig:Multimode}\textbf{a} and Fig. \ref{fig:Multimode}\textbf{f} respectively. Each microring is coupled with a straight bus waveguide in a conventional single-point coupling scheme. For the smaller 200 $\mu$m microrings, the width of the bus waveguide is varied from 1600 to 2400 nm, with a gap ranging between 650 and 750 nm, allowing for the cavities to be swept from undercoupling into overcoupling conditions. The larger 400 $\mu$m microrings have constant bus waveguide widths of 2000 nm and gaps ranging from 650 to 800 nm. Thanks to the negligible bending loss experienced by the TE-modes, the larger 400 $\mu$m radius is not necessary to support ultra-low-loss operation, as shown in Fig. \ref{fig:sim}\textbf{f}. For TE-polarization, the smaller type-1 microrings exhibit a consistent, high most-probable intrinsic-Q of $3.4\times10^6$ across 8 tested resonators between wavelengths of 1490 nm and 1630 nm (Fig \ref{fig:Multimode}\textbf{b}), including the region of hydrogen absorption near 1520 nm. A peak intrinsic-Q of $Q_i=4.6\times10^6$ is extracted for the smaller type-1 microrings at 1605 nm (Fig. \ref{fig:Multimode}\textbf{c}), suggesting to an ultra-low propagation loss of 7.5 dB/m. To our knowledge, these numbers represent the highest quality factor achieved at telecom and the lowest loss attained at any wavelength in Si$_3$N$_4$-on-sapphire to-date, and the first demonstration of loss below the 0.1 dB/cm (10 dB/m) mark for the platform. Notably, the high-Q resonances shown in \ref{fig:Multimode}\textbf{d} are no outliers. Figures \ref{fig:Multimode}\textbf{d} and \textbf{e} clearly demonstrate that the best performance type-1 resonator exhibits a mean intrinsic-Q of $3.6\times10^6$ between 1490 nm and 1630 nm, corresponding to an average propagation loss just above the 10 dB/m mark. Inspecting Fig. \ref{fig:Multimode}\textbf{e} for wavelengths above 1550 nm, the majority of resonances fall below 10 dB/m in propagation loss. The trends for the characteristic increase in loss (and associated decrease in $Q_i$) in Figs. \ref{fig:Multimode}\textbf{d} and \textbf{e} are attributed to hydrogen absorption near 1520 nm, suggesting that higher temperature annealing (1200 $^\circ$C) would unlock even superior performance in this spectral region, as corroborated in the lowest loss implementations of traditional Si$_3$N$_4$ on SiO$_2$ PICs \cite{LipsonSiN67MQ17}.

Figure \ref{fig:Multimode} illustrates the performance of TM-polarized modes for the larger type-2 microrings. Across 4 devices measured from 1490 nm to 1630 nm, the distribution of intrinsic-Q shown in Fig. \ref{fig:Multimode}\textbf{g} indicate a most-probable $Q_i = 2.9\times10^6$, slightly lower than that of the TE-polarized modes as expected, and peak intrinsic quality factors of $3.6\times10^6$ past 1600 nm shown in Fig. \ref{fig:Multimode}\textbf{h}, corresponding to sub 10 dB/m losses in the regime of ultra-low loss. In Figs. \ref{fig:Multimode}\textbf{i} and \textbf{j}, the highest performance type-2 microring follows a different trend for its loss and quality factor than the type-1 microring shown in Figs. \ref{fig:Multimode}\textbf{d} and \textbf{e}, as TM-polarized modes, with their reduced confinement, exhibit stronger interactions with the etched waveguide sidewall than TE-polarized modes, resulting in them being more susceptible to Rayleigh-type scattering loss at waveguide edge. This is clearly shown in the continued upward trend of intrinsic-Q for the TM modes as wavelength increases in comparison to the overturn of the initial upward trend of the TE modes toward long wavelengths due to, hypothetically, absorption loss, suggesting the existence of impurities in the Si$_3$N$_4$ layer.

\subsection{Characterization of Photonic Molecules}

The generation of frequency combs in normal dispersion Si$_3$N$_4$ waveguides requires perturbation of the dispersion to create a region of local anomalous dispersion. In this regard, strongly coupled double microring resonators, i.e., the photonic molecules, have been introduced to reliably engineer local anomalous dispersion and generate dark-pulse and bright-pulse solitons at high conversion efficiency \cite{helgason_surpassing_2023}. In this work, we implement photonic molecules, illustrated in Fig. \ref{fig:Coupledring}\textbf{a}, through two strong-coupled single-mode microrings with widths of 2.2 $\mu$m and radii of 200 and 190 $\mu$m  for the main and auxiliary cavities respectively. This choice of radii corresponds to approximately 20 modes between split modes. The coupling rate between the two cavities is controlled by the gap between the microrings, varied between 500 and 650 nm, while the precise alignment of the resonances of the two cavities is controlled by a microheater on the auxiliary cavity.

Letting $f_a$ and $f_b$ be the bare-cavity resonances of the main (a) and auxiliary (b) microrings in the un-coupled basis, usage of the microheater allows for direct tuning of the frequency difference $f_b - f_a$, facilitating measurements of the coupling rate between the two cavities. To measure the coupling rate ($\kappa$) of the photonic molecules, we measure avoided mode-crossings of symmetric and anti-symmetric modes $f_\pm$ by performing short-range (0.2 nm) laser sweeps around the split resonances at successively higher applied heater currents. Equation (\ref{eq:splitting}), shown below, describes the relationship between the split resonance frequencies, the placement of the two cavity resonances in the uncoupled basis, and the coupling rate between the two cavities \cite{helgason2021dsin}.
\begin{equation}
\label{eq:splitting}
    f_{\pm} = \frac{1}{2}(f_a + f_b) \pm\sqrt{\frac{1}{4}(f_a - f_b)^2 + \left( \frac{\kappa}{2\pi}\right)^2}
\end{equation}
Figure \ref{fig:Coupledring}\textbf{b} shows a typical trend of the raw measurement, with avoided mode crossing behavior visible, but contributing less to the overall trend than the clear quadratic common-mode movement of the split resonances, owing to frequency shift that is proportional to the heater power and thereby quadratic to the current, i.e., $\Delta f \propto \Delta n \propto T\propto P_{\rm heater} \propto I_{\rm heater}^2$. By employing a fiber-loop resonator reference cavity with a near-constant free spectral range (FSR), the obstacles of sweep speed jitter and uncertain starting wavelength are circumvented. With common-mode movement $(f_+ + f_-)/2$ removed, Fig. \ref{fig:Coupledring}\textbf{c} clearly illustrates the avoided mode crossing. Zooming in further, Fig. \ref{fig:Coupledring}\textbf{d} shows the coupling rate inferred from the minimum difference between the split resonances ($f_+ - f_- = 2\kappa$), occurring at $f_a = f_b$, when the two cavity resonances in the uncoupled basis perfectly align. Figure \ref{fig:Coupledring}\textbf{e} further depicts the coupling rates for multiple photonic molecules with different main-auxiliary microring coupling gaps near 1575 nm, demonstrating good agreement between experiment and simulation. The tunability of the coupling rate is essential as it, combined with the inherent dispersion of the waveguide, controls the effective maximum region of the local anomalous dispersion, thereby controlling nonlinear gain as a result. 

\subsection{Generation of Local Anomalous Dispersion}
\label{sec:singlemodeperf}
\begin{figure}
\centering\includegraphics[width=1\linewidth]{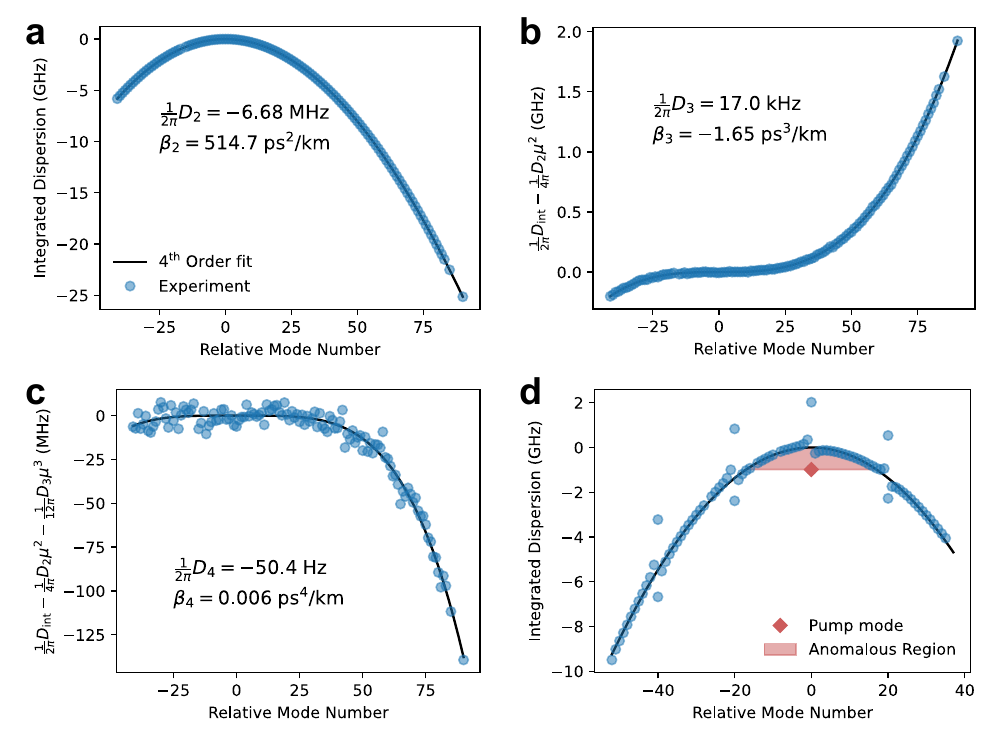}
\caption{{\bf Dispersion Measurements of Single-Mode and Coupled Rings.} Measured integrated dispersion curves for a single-mode un-coupled ring from 1490 nm to 1630 nm for TE-polarization with clear second (\textbf{a}), third (\textbf{b}), and fourth (\textbf{c}) order behavior extracted. \textbf{d} Measured integrated dispersion for a coupled-ring photonic molecule from 1540 nm to 1630 nm, with generated region of local anomalous dispersion highlighted. For all integrated dispersion curves, $\mu = 0$ is near 1575 nm.}
\label{fig:Dispersion}
\end{figure}

The three-way connection of the coupling-strength between the two cavities, the un-coupled-cavity dispersion, and region of local anomalous dispersion generated becomes most evident when observing the integrated dispersion profile of the coupled rings. Integrated dispersion ($D_{\rm int}$) describes the walk-off of cavity resonances away from the constant FSR approximation. Defining integrated dispersion of the mode $\mu$ relative to an arbitrary pump $\omega_0$, the profile of a un-coupled ring can be described by $D_{\rm int}(\mu) = (\omega_\mu - \omega_0) - D_1\mu = \frac{1}{2}D_2\mu^2 + \frac{1}{6}D_3\mu^3 + \frac{1}{24}D_4\mu^4 + \cdots$, where $D_1$ is the constant-FSR approximation and the higher order terms, $D_i$ for $i\geq 2$, account for the walk-off. The short-range walk-off is largely quadratic, originating from group velocity dispersion ($\beta_2$); however, higher order terms are still present and become more evident with greater distance from the pump. From this frame of reference, anomalous dispersion, required for nonlinear gain, is achieved when $D_{\rm int}(\mu) > 0$. 

Figures \ref{fig:Dispersion}\textbf{a}, \textbf{b}, and \textbf{c} show the extracted integrated dispersion of a single-mode, un-coupled 200 $\mu$m radius microring, with a 2.2 $\mu$m width, which was used as a dispersion reference for the coupled cavities. To extract integrated dispersion terms up to the fourth order, we perform a broad transmission measurement from 1490 to 1630 nm with the addition of a fiber-loop reference cavity to correct for jitter in the laser sweep. Each of the extracted higher order integrated dispersion terms $\frac{1}{2\pi}D_2 = -6.68 \rm\; MHz$, $\frac{1}{2\pi}D_3 = 17 \rm\; kHz$, and $\frac{1}{2\pi}D_4 = -50.4 \rm\; Hz$ correspond to higher order propagation constants $\beta_2 = 514.7 \;\rm ps^2/km$, $\beta_3 = -1.65 \;\rm ps^3/km$, and $\beta_4 = 0.006 \;\rm ps^4/km$, calculated from the measured integrated dispersion terms and the assumed cavity length. These experimental values show good agreement with the COMSOL-simulated dispersion values of  $\beta_2^{\rm sim} = 513.2 \;\rm ps^2/km$, $\beta_3^{\rm sim} = -1.72 \;\rm ps^3/km$, and $\beta_4^{\rm sim} = 0.010 \;\rm ps^4/km$, indicating the validity of the measurement and the precision of fabrication. Discrepancies at this scale likely originate from minute imperfections in fabrication, especially in the uncertainty in waveguide width and height. From the extracted dispersion values and the overall trend in Fig. \ref{fig:Dispersion}\textbf{a}, it is clear, as expected, that a lone single-mode ring does not exhibit anomalous dispersion. However, the split modes generated by the photonic-molecule perturb the profile in a manner that creates a region of local anomalous dispersion, permitting nonlinear gain and frequency comb generation.

In the case of the photonic molecule, the anti-symmetric mode, $f_-$, finds itself shifted downwards from the un-coupled integrated dispersion curve by an amount of $2\pi(f_{a}- f_-)$. This shift is visible in Fig. \ref{fig:Dispersion}\textbf{d}, where strong splittings occur every 20 modes. When the resonances of the un-coupled rings are perfectly aligned, $f_a = f_b$, this shift is exactly the coupling rate between the two cavities, $\kappa$. From the perspective of a resonant pump placed at $f_-$, this yields a region of local anomalous dispersion, encompassing mode numbers $|\mu| \lesssim \sqrt{{2\kappa}/{D_2}}$. Away from perfect alignment, the region of local anomalous dispersion changes, highlighting cavity alignment, coupling rate, and dispersion as important tuning knobs for frequency comb generation. While the last two are set for a given fabricated device, the first can be actively manipulated, offering intriguing opportunities for in-device dispersion tailoring, further explored in the following section.

\subsection{Normal Dispersion Frequency Comb Generation}
\begin{figure*}
\centering\includegraphics[width=1\linewidth]{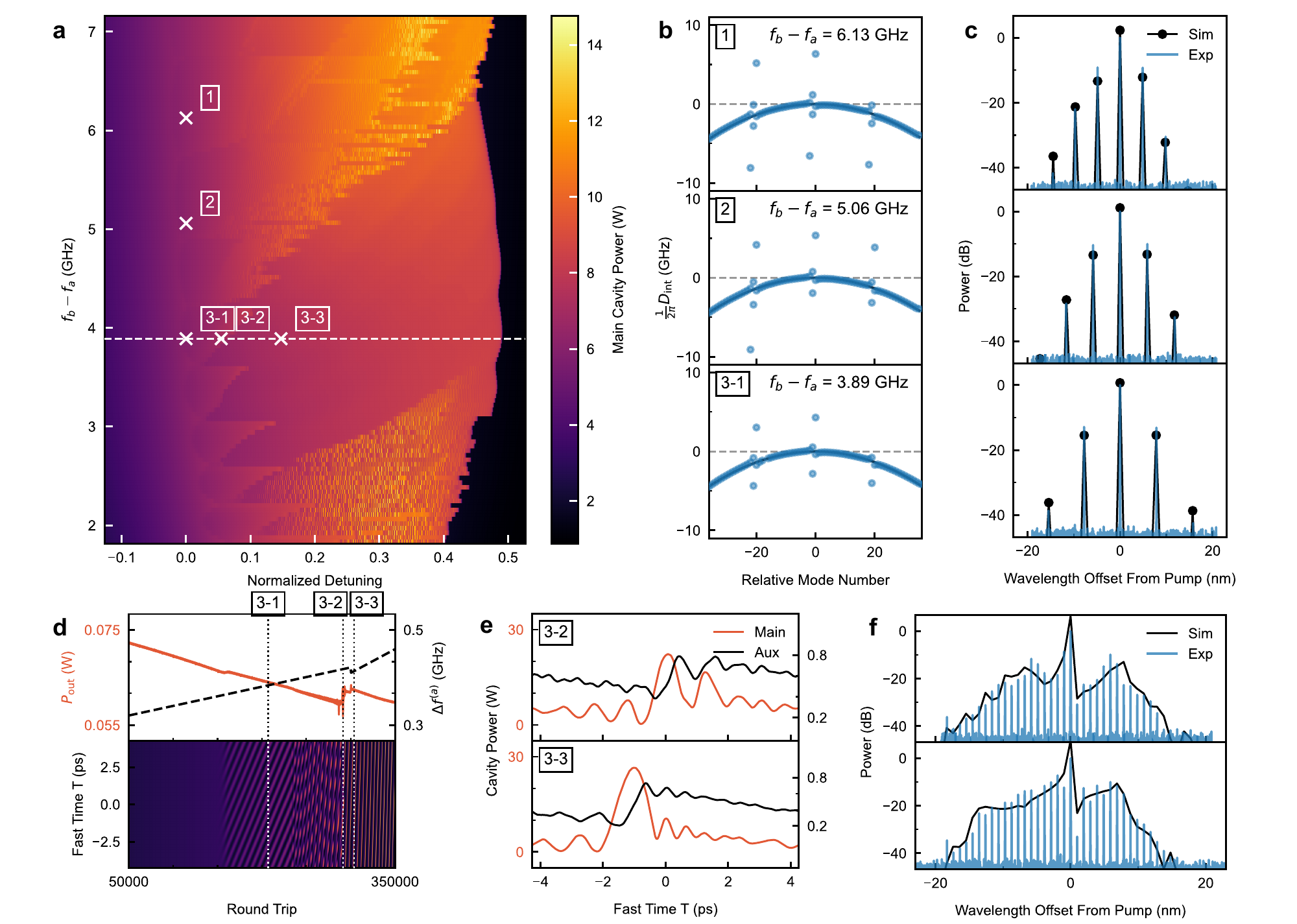}
\caption{\textbf{Splitting-Dependent Dissipative Soliton Dynamics in the Photonic Molecule.}
\textbf{a,} Heatmap of the main-cavity intracavity power as a function of normalized detuning $u$ and mode splitting $f_b-f_a$, at a fixed on-chip pump power $P_{\mathrm{in}}=\SI{90}{\milli\watt}$. Markers 1, 2, and 3-1 indicate three distinct Turing-roll states, whose dispersion profiles and spectra are shown in (\textbf{b}) and (\textbf{c}) ($f_b-f_a\approx6.13$, \SI{5.06}{}, and \SI{3.89}{\giga\hertz}, respectively). The detuning scan along the row containing state 3-1 is presented in (\textbf{d}), with its corresponding chaotic (3-2) and single-soliton (3-3) states detailed in (\textbf{e}) and (\textbf{f}).
\textbf{b,} Measured integrated dispersion $D_{\mathrm{int}}$ versus relative mode number $\mu$ at splittings 1, 2, and 3-1. The quadratic fits (black curves) yield $D_2/2\pi\approx-6.81$, $-6.79$, and \SI{-6.82}{\mega\hertz}, confirming normal dispersion across all three states.
\textbf{c,} Simulated (black) and measured (blue) Turing-roll spectra at states 1, 2, and 3-1, exhibiting comb-line spacings equal to 5, 6, and 8 cavity free spectral ranges (FSRs), respectively.
\textbf{d,} Bus-waveguide output power (left axis) and main-cavity detuning $\Delta f^{(a)}$ (right axis, dashed curve) versus round-trip number during the detuning scan along the row containing states 3-1, 3-2, and 3-3 in (\textbf{a}). The bottom panel displays the corresponding main-cavity intracavity intensity evolution over round trips and fast time. Vertical lines indicate the specific round trips corresponding to the chaotic (3-2) and single-soliton (3-3) states.
\textbf{e,} Simulated intracavity power versus fast time in the main (orange) and auxiliary (black) cavities for both the chaotic and single-soliton states.
\textbf{f,} Simulated (black) and measured (blue) output spectra corresponding to the chaotic and single-soliton states.}
\label{fig:NLO}
\end{figure*}

Previous studies have demonstrated the generation of normal-dispersion dissipative Kerr solitons (DKSs) in coupled-cavity architectures \cite{yuansoliton2023}, featuring robust operation, deterministic soliton triggering \cite{rebolledo2023platicon}, high conversion efficiency \cite{helgason_surpassing_2023}, and intrinsic thermal self-cooling \cite{nishimoto2025selfcooling}. Adopting this approach, robust generation of DKSs on our Si$_3$N$_4$-on-sapphire platform is achieved, positioning the platform as a versatile and powerful tool for future applications for spectroscopy and precision frequency metrology.

In the experiment to generate DKSs, an amplified tunable continuous-wave laser is coupled into a photonic molecule on the $\text{Si}_3\text{N}_4$-on-sapphire chip with a $500\ \text{nm}$ bus-to-waveguide gap and a $575\ \text{nm}$ microring-to-microring gap mounted on a stage without active temperature control. The transmitted output is split for simultaneous real-time monitoring of power transmission and optical spectra. The pump laser frequency is controlled by an arbitrary waveform generator (AWG).

The thorough investigation of the comb generation dynamics on this platform entails numerical simulations to explore the parameter space followed by experimental validation of the numerical predictions. Figure~\ref{fig:NLO}\textbf{a} presents a heatmap of the main-cavity intracavity power as a function of normalized detuning $u$ and mode splitting $f_b-f_a$. During the detuning scan, the system consistently transitions first into a Turing-roll state. Because the mode splitting directly modifies the local dispersion profile (Fig.~\ref{fig:NLO}\textbf{b}), it inherently alters the parametric gain landscape. Consequently, the spacing of the newly generated Turing rolls---defined by the position of the primary sidebands---depends fundamentally on the splitting magnitude. As demonstrated by the measured spectra in Fig.~\ref{fig:NLO}\textbf{c}, a larger splitting corresponds to a smaller Turing-roll order; specifically, splittings of \SI{6.13}{}, \SI{5.06}{}, and \SI{3.89}{\giga\hertz} yield Turing rolls with spacings of 5, 6, and 8 cavity free spectral ranges (FSRs), respectively, showing excellent agreement with our numerical predictions. Furthermore, the accessible detuning range that sustains these Turing rolls varies with the splitting, outlining the distinct triangular boundary regions clearly visible in the heatmap.

Beyond the initial Turing rolls, the numerical simulations reveal that the spatiotemporal evolution of the optical field exhibits a profound dependence on the mode-splitting regime. In regions with relatively small splitting, the primary Turing-roll state rapidly devolves into an unstable, chaotic comb state, which is also confirmed by the spectral measurements. In contrast, other optimal splitting regimes bypass this extended instability and facilitate a direct transition into a stable single-soliton state. Interestingly, under conditions of much larger splitting, the numerical simulations reveal the emergence of dual-peak, symmetric temporal structures, which closely resemble the dynamics characteristic of soliton crystals. More details of Fig. \ref{fig:NLO}\textbf{a} are provided in the Supplementary Information..

Guided by these numerical insights, we experimentally target the optimal splitting regimes where single dissipative solitons can be deterministically generated, corresponding to the extensive, uniformly colored areas in the heatmap of Fig.~\ref{fig:NLO}\textbf{a}. Figure~\ref{fig:NLO}\textbf{d} illustrates the continuous evolution of the intracavity field alongside the effective detuning during a cavity scan (along the row containing state 3-1). Notably, a distinct detuning kickback upon reaching the soliton state is manually introduced. In the actual experiment, the thermal accumulation rate within the cavity cannot keep pace with the laser scanning speed, leaving the system in a non-equilibrium thermal state during the sweep. Once the experimental scan stops at the soliton state, the delayed heat accumulation catches up and causes the cavity temperature to rise. This continuous temperature increase shifts the cavity resonance, which is physically equivalent to a decrease in detuning. The intracavity intensity follows a well-defined route: it begins from a continuous-wave (CW) background, evolves into Turing rolls, transitions through an unstable chaotic comb, temporarily forms a multi-pulse structure, and finally collapses into a stable single DKS. The corresponding simulated time-domain optical field of this stable single pulse is shown in Fig.~\ref{fig:NLO}\textbf{e} and Fig.~\ref{fig:NLO}\textbf{f} demonstrate the excellent agreement between the simulated and experimentally measured output spectra for both the chaotic and single-soliton states.

\section{Discussion}

We have identified multiple avenues to further reduce the intrinsic loss of the Si$_3$N$_4$-on-sapphire platform. First, the impurities in Si$_3$N$_4$ are speculated to stem from the usage of a less-restrictive class of furnace for the LPCVD Si$_3$N$_4$ layer, in compliance with the current internal material segregation requirements of the shared cleanroom facility. With deposition of Si$_3$N$_4$ from a more restrictive LPCVD furnace and higher temperature annealing, the performance is expected to be lifted across the entire wavelength range of interest. Furthermore, the width of these multimode waveguides is kept conservative to suppress the excitation of higher-order modes and maintain quasi-single mode operation. Due to the reduced index contrast of sapphire substrates, 3-$\mu$m-wide waveguides that support merely 3 modes beyond 1550 nm (Fig. \ref{fig:sim}\textbf{b}) are not nearly as multi-modal as their traditional Si$_3$N$_4$-on-SiO$_2$ counterparts of similar thickness. Since the width of the waveguide correlates strongly with sidewall interaction that contributes to scattering loss, implementation of cavities with wider waveguides and adiabatic bends to suppress higher-order mode excitation will further suppress scattering loss and increase the intrinsic-Q accessible on this platform \cite{ThickSi3N4MultiModeLipson21}. 

In comparison to existing platforms, Si$_3$N$_4$-on-sapphire possesses an array of unique advantages. By integrating Si$_3$N$_4$ with sapphire substrates, the platform exhibits farther spectral reach than traditional Si$_3$N$_4$ on SiO$_2$ photonic integrated circuits while maintaining better short-wavelength and telecom performance than other MIR-compatible platforms such as Si-on-insulator and Si-on-Sapphire, providing maximal spectral coverage while minimizing loss across its accessible regime. In our fabricated devices, the 3-$\mu$m thick SiO$_2$ cladding layer is chosen to keep fabrication in-line with our conventional Si$_3$N$_4$ on SiO$_2$-on-Si photonic integrated circuits; however, for applications in the mid-IR, this cladding layer would need to be swapped for a material with better long wavelength transmission, such as amorphous Al$_2$O$_3$. Moreover, the sapphire substrate supports direct low-stress deposition of high-quality thin and thick-film Si$_3$N$_4$, lowering the fabrication cost and complexity of both low-confinement ultra-low-loss and high-confinement dispersion engineered waveguides. These combined benefits of ultra-low loss, accessible nonlinearity, wide spectral coverage, and reduced fabrication complexity will motivate integrated photonics community to further explore this rising platform for nonlinear and quantum integrated photonics.

In summary, we have successfully demonstrated the design and fabrication of ultra-low-loss photonic integrated circuits on the Si$_3$N$_4$-on-sapphire platform, achieving, for the first time, sub-0.1 dB/cm losses for both TE and TM polarization, losses as low as 7.5 dB/m, and intrinsic quality factors as high as $4.6\times10^6$. Continued development based on these benchmark numbers is poised to open compelling pathways in both fabrication and design to achieve even higher performance. This work not only showcases ultra-low loss on the Si$_3$N$_4$-on-sapphire platform, but also presents robust device structures to access a rich landscape of nonlinear states of light, highlighting the deterministic generation of dissipative solitons. Tied together, these results underscore the tremendous potential of the Si$_3$N$_4$-on-sapphire platform for scalable and robust integrated nonlinear and quantum photonics.

\section{Methods}
\subsection{Fabrication of Si$_3$N$_4$-on-Sapphire Photonic Integrated Circuits}
\label{sec:FabMethods}

The fabrication of Si$_3$N$_4$-on-sapphire PICs, as detailed in Fig. \ref{fig:fab}\textbf{a}, begins with a 450 $\mu$m thick, single-side polished, epi-ready, c-plane sapphire wafer. After RCA cleaning, a 500 nm thick layer of stoichiometric Si$_3$N$_4$ is deposited by LPCVD. Adopting the etching method first shown by Liu {\em et al.} in Ref. \citenum{ShuaiACS25}, the Si$_3$N$_4$ layer is then cladded by a 700-nm-thick sacrificial a-Si hardmask layer using plasma-enhanced chemical vapor deposition (PECVD). Following oxygen (O$_2$)-plasma treatment and HMDS vapor prime, a coating of ma-N 2405 negative electron-beam resist is spun on the sample surface. The waveguide patterns are then precisely defined using electron-beam lithography with a 2 nA current, an 8 nm step size, and a dose of 700 $\mu$C/cm$^2$. The exposed resist is then developed by MF-319, a tetra-methyl ammonium hydroxide (TMAH) based developer, before thermal reflow to smoothen the defined patterns. Inductively-coupled plasma reactive ion etching (ICP-RIE) in a Hyrdogen Bromide (HBr) and Helium (He) environment transfers the pattern to the a-Si layer. After removing the remnant ma-N resist via O$_2$-plasma etching, the Si$_3$N$_4$ is etched via ICP-RIE in a C$\rm _x$F$\rm _y$ environment, revealing a smooth, nearly vertical sidewall, as visible in Fig. \ref{fig:fab}\textbf{f}. The remaining a-Si hardmask layer is etched away using isotropic Xenon Diflouride (XeF$_2$) etching. With the Si$_3$N$_4$ waveguide layer fully patterned and exposed, the device receives its first high temperature nitrogen (N$_2$) anneal at 1100 $^\circ$C for 6 hours, to drive out hydrogen in the Si$_3$N$_4$ layer from remnant N--H and Si--H bonds. Once removed from the annealing furnace, the device is RCA cleaned before receiving 3 $\mu$m of PECVD SiO$_2$ cladding and a final 1100 $^\circ$C N$_2$ anneal, completing the photonic layer. To control the splitting of the photonic molecules, a metal layer on top of the photonic layer is necessary to accommodate microheaters, fabricated by patterning an LOR-10B/S1813 layer of photoresist via UV stepper photolithography, depositing a 250 nm thick layer of platinum (Pt) after AZ-726 development, and performing lift-off via Remover PG. Edge couplers are defined by dicing using a diamond blade and polishing the facet using diamond polishing paper with successfully smaller grits descending from 60 to 0.25 $\mu$m. After dicing and polishing, the chip is cleaned in a heated bath of PRS-2000 solvent.   

\subsection{Transmission Characterization and Loss Extraction}
In the experimental setup, a tunable semiconductor laser (Santec TSL-770) is fed into a fiber polarization controller (FPC) before being coupled onto the chip via a tapered lensed fiber to preferentially excite fundamental TE or TM modes of the device with high efficiency. The light is coupled off-chip using another tapered lensed fiber before being detected by a low-noise photodetector (Newport 1811). To measure the transmission spectrum, the laser is swept across its usable range (1490 nm to 1630 nm) at a rate of 20 nm/s to avoid modification of the Lorentzian resonance line shape due to thermo-optic effects. The resonance dips of the normalized sweep data are then identified and fit to a Lorentzian transmission function:
\begin{equation}
\label{eq:LorentzTrans}
T(\delta\lambda) = \left|
\frac{
\left(\frac{\kappa_i}{2}\right)^2
-\left(\frac{\kappa_{e}}{2}\right)^2
-(\delta\lambda)^2
+i\,\delta\lambda\,\kappa_i
}{
\left(\frac{\kappa_i}{2}+\frac{\kappa_{e}}{2}\right)^2
-(\delta\lambda)^2
+i\,\delta\lambda\,(\kappa_i+\kappa_{e})
}
\right|^2,
\end{equation}
where $\delta\lambda$ is the wavelength detuning from resonance \cite{ShuaiACS25}.

After fitting the intrinsic and extrinsic linewidths ($\kappa_i, \; \kappa_e$), intrinsic and extrinsic quality factors are calculated using $Q_{i, e} = {\lambda_0}/{\kappa_{i, e}}$, where $\lambda_0$ is the resonance wavelength. From the intrinsic quality factor $Q_i$, the propagation loss can then be calculated using
\begin{equation}
\label{eq:LossFromQ}
    \alpha = \frac{2\pi n_g}{\lambda_0 Q_i} = \frac{2\pi\lambda_0}{\Delta\lambda_{\rm FSR}LQ_i},
\end{equation}
where the group index ($n_g$) can either be found from simulation or calculated using the FSR and length ($L$) of the cavity \cite{ThickSi3N4MethOverStress2013}.

\subsection{Avoided Mode Crossing and Dispersion Measurements}
Building off of the same experimental setup as utilized for loss extraction, we include an additional fiber-loop reference cavity that is simultaneously swept during the measurement. The fiber-loop resonator has an FSR of $\Delta f_{\rm FSR} = 93.66$ MHz, which was calibrated by an electro-optic modulator (EOM), where the drive frequency was increased until the first sidebands overlapped. To account for the time-of-flight difference between the two paths, the sweep speed of laser is reduced to 2 nm/s for measurements that require the fiber-loop resonator. 

Measurements of the avoided mode crossings were conducted near 1575 nm and consisted of short-range 0.2 nm sweeps at 2 nm/s around the split-resonance modes. After each sweep, the current applied to the auxiliary ring microheater was increased in increments of 0.5 mA. A measurement span of 40 mA was chosen to ensure full coverage of the avoided mode crossing. The recorded transmission spectra were then converted from a time-basis to a frequency-basis according to the fiber-loop resonator FSR. The mid-point between the two split modes $(f_+ + f_-)/2$ was set as the reference point from which the difference in frequency between the two split modes was tracked. The minimum experimental value of the difference $f_+ - f_-$ was used to extract the coupling rate $\kappa = \min(f_+-f_-)/2$. 

Dispersion measurements were carried out using the same experimental setup, but employed broader sweeps starting from 1490 or 1540 nm and ending at 1630 nm. Lorentzian fitting of the transmission spectrum determined the resonant wavelength of the modes, which were then fed into our dispersion fitting code. Choosing one of the cavity modes near 1575 nm as the central mode $\mu = 0$, we measure the frequency difference $f_\mu - f_0$ using the reference fiber-loop resonator. This difference is then iteratively fit to a fourth-order model, permitting direct extraction of the integrated dispersion parameters $D_1$ through $D_4$. From these values and the assumed cavity length, higher-order propagation constants $\beta_1$ through $\beta_4$ are calculated. Full fourth-order fitting was exclusively performed on the un-coupled reference single-mode rings, as the strong-coupling between the two microrings in the photonic molecules obscures any higher-order behavior beyond the quadratic term, $D_2$. As a result, photonic-molecule dispersion measurements, as shown in Figs. \ref{fig:Dispersion}\textbf{d} and \ref{fig:NLO}\textbf{b}, are only fitted up to the second order.

\subsection{Details of Comb Generation Experiment}
In the comb generation setup, a tunable continuous-wave laser (Santec TSL-770) passing through an optical isolator is amplified by an erbium-doped fiber amplifier (EDFA). Its output polarization is optimized via a polarization controller before coupling into the $\text{Si}_3\text{N}_4$-on-sapphire chip using lensed fibers. The chip is mounted on a stage without active temperature control. The transmitted signal is routed through optical splitters and simultaneously sent to an optical spectrum analyzer (OSA), a power meter, and a fast photodetector connected to an oscilloscope for real-time spectral and temporal monitoring. Laser frequency modulation is driven by an arbitrary waveform generator (AWG) connected to the laser's fine-tuning voltage port (with a modulation sensitivity of $10\ \text{GHz/V}$). To deterministically generate a single soliton state, the pump laser is initially swept across the resonance at a fast rate of $\sim 380\ \text{GHz/s}$. Upon triggering the targeted comb state, the pump laser is immediately switched to an ultra-slow frequency sweep rate of $\sim 19\ \text{kHz/s}$.

\bigskip

\backmatter

\bmhead{Author contributions} A. Al-Hallak, S. Liu, J. Liu, Z. Zhang, and Z. Mi conceived the idea. Si$_3$N$_4$ on sapphire wafers were prepared by S. Liu and J. Liu, with chip fabrication performed by A. Al-Hallak. FEM simulations were carried out by A. Al-Hallak and M. Sarram, with S. Chen and J. Hu assisting in FDTD simulations of the coupled-ring photonic molecules. A. Al-Hallak and S. Liu designed the chip. Low-power characterization and analysis for quality factor and dispersion extraction was performed by A. Al-Hallak, J. Hu, and R. Yusuf. S. Liu, A. Al-Hallak, and Y. Lang developed the dispersion measurement and analysis techniques. Avoided mode-crossing measurements were devised by A. Al-Hallak and S. Chen, performed by S. Chen and C. Rodriguez, and analyzed by A. Al-Hallak and C. Rodriguez. K. Zhou developed numerical simulation methods to predict soliton states and devised the high-power experiments for frequency comb generation. Measurements of frequency comb and soliton states were carried out by K. Zhou and A. Al-Hallak. The manuscript was written by A. Al-Hallak, K. Zhou, Z. Zhang, and Z. Mi, with commentary and assistance from all authors. 

\bmhead{Conflict of interest}
The authors declare that a provisional patent application (010109-26023P-US) has been filed for the devices laid out in this work.

\bmhead{Supplementary information} Supplementary materials for this work will be provided upon publication. 

\bmhead{Acknowledgments} We acknowledge funding support from the National Science Foundation Grant No. 2317471 and 2609754, the Army Research Office Grant No. W911NF2420210, and the University of Michigan. Device fabrication was carried out on-site in the Lurie Nanofabrication Facility (LNF).

\bibliography{SiN_sapphire.bib}

\end{document}